\documentclass[5p,times]{elsarticle}

\usepackage{graphicx}  
\usepackage{subfigure}
\usepackage{multirow}
\usepackage{float}
\usepackage{tikz}
\usepackage{xfrac}
\usepackage{comment}
\usepackage{dirtytalk}
\include{tikzspecs}
\include{booktabs}
\usepackage{blindtext}
\usepackage{cuted}

\usepackage{hyperref}
\hypersetup{
colorlinks=true,
urlcolor= blue,
citecolor=blue,
linkcolor= blue,
}

\usepackage{fancyhdr}
\usepackage{longtable}
\usepackage{parskip}
\usepackage[T1]{fontenc}
\usepackage{dcolumn}   

\usepackage{bm}        
\usepackage{amsfonts}  
\usepackage{amsmath}   
\usepackage{amssymb}   
\usepackage{booktabs}
\usepackage{cleveref}
\usepackage{balance}
\crefformat{figure}{#2{\color{black}#1}#3}
\crefformat{table}{#2{\color{black}#1}#3}

\newcommand{\pwiseout}{\end{array}\right.}

\newcommand{\Plus}[0]{\texttt{+}}

\setcitestyle{number,open = {[},close={]}}

\begin{document}
\title{Flavour and Energy Dependence of Chemical Freeze-out Temperatures in Relativistic Heavy Ion Collisions from RHIC-BES to LHC Energies}
\author{Fernando Antonio Flor}
\ead{faflor@uh.edu}
\author{Gabrielle Olinger}
\author{Ren\'e Bellwied}
\address{Department of Physics, University of Houston, Houston, Texas 77204, USA}
\date{\today}

\begin{abstract}
We present calculations of the chemical freeze-out temperature ($T_\mathrm{ch}$) based on particle yields from STAR and ALICE measured at collision energies ranging from $\sqrt{s_{\rm{NN}}} = $ 11.5 GeV to 5.02 TeV.  Employing the Grand Canonical Ensemble approach using the Thermal-FIST Hadron Resonance Gas model package, we show evidence for a flavour-dependent chemical freeze-out in the crossover region of the QCD phase diagram. At a vanishing baryochemical potential, we calculate light and strange flavour freeze-out temperatures $T_\mathrm{L} = 150.2 \pm 2.6$ MeV and  $T_\mathrm{S} = 165.1 \pm 2.7$ MeV, respectively.
\end{abstract}


\maketitle
\setlength{\parskip}{0em}

\section{Introduction}

\indent In depth determinations of a pseudo-critical temperature based on continuum extrapolations of temperature dependence of the chiral susceptibilities on the lattice, in comparison to calculations using Statistical Hadronization Models (SHM) using particle yields from experiments at the Relativistic Heavy Ion Collider (RHIC) and the Large Hadron Collider (LHC), indicate that chemical freeze-out and hadronization coincide near the phase boundary in the Quantum Chromodynamics (QCD) phase diagram \cite{Lattice, WB1, WB2, HotQCD1, Stachel_2014}. Whether this transition from quark to hadron degrees of freedom occurs at a uniform temperature for all quark flavours remains a question of interest. \\
\indent SHMs have been successful in adequately reproducing hadronic particle abundances over nine orders of magnitude in high energy collisions of heavy ions over a wide range in energy \cite{Cleymans1998,PBMNature}. In these calculations, assuming a thermally equilibrated system, experimental particle yields in relativistic heavy ion collisions serve as an anchor for the determination of common freeze-out parameters in the QCD phase diagram -- namely, the baryo-chemical potential ($\mu_\mathrm{B}$) and the chemical freeze-out temperature ($T_\mathrm{ch}$). The resulting parameters can also be compared with independently obtained results from either lattice QCD based susceptibility calculations of conserved quantum numbers or measurements of higher order fluctuations of net-particle distributions. 

\section{Sequential Flavour Freeze-out}
\indent Continuum extrapolated susceptibility calculations of single flavour quantum numbers on the lattice \cite{Ratti2012, SuscRat} have shown a difference in the determined freeze-out temperatures between flavours in the crossover region of the QCD phase diagram. This effect is likely due to the difference in the bare quark masses which is not negligible in a thermally equilibrated deconfined system near the phase boundary. In particular, a comparison of the flavour specific susceptibility ratios ${\chi_{4}}/{\chi_{2}}$, suggested as an observable for directly determining freeze-out temperatures \cite{Karsch2012}, show a deviation of the lattice and Hadron Resonance Gas (HRG) model calculations coinciding at the peaks of the lattice data, which occur at flavour specific temperatures differing by $15 - 20$ MeV from light to strange quarks \cite{SuscRat}. Net-particle fluctuation measurements by the STAR collaboration have also shown comparable temperature differences between the light and strange mesons \cite{Adamczyk:2013dal, NetK, BluhmKaon}. 

\indent Thermal fits to experimental data via SHMs have shown similar results, depicting a difference in the freeze-out temperatures between flavours in the crossover region. The STAR Collaboration  recently published the dependence of their thermal fits to the yields on the particle species included in the fit rendering a freeze-out temperature about $10 - 15$ MeV lower when fitting only pions, kaons and protons than a common freeze-out temperature extracted from fits to yields of all measured particle species \cite{BES}. Moreover, it has been shown that assuming two distinct freeze-out temperatures improves the overall fit to ALICE data \cite{Chatterjee_2017}. Thus, a point of interest arises when comparing the extracted freeze-out parameters obtained using different sets of particles in the SHM calculation. This approach was recently performed in an HRG-based study on both diagonal and non-diagonal second order correlators of conserved charges \cite{Bellwied_PRD_2020}.

\section{Model and Data Preparation}
The entirety of our analysis was performed using the open source Thermal FIST (The FIST) thermal model package \cite{FIST}. Without loss of generality, The FIST is a user-friendly package within the family of HRG Models. Although there exists a wide range of options for the HRG model within The FIST framework, we restricted our analysis to the default -- namely, modeling an ideal non-interacting gas of hadrons and resonances within a Grand Canonical Ensemble (GCE). 

All our calculations used the PDG2016\Plus{} hadronic spectrum \cite{PDG2016+} as the HRG input list, including a total of 738 states (i.e. *, **, *** and **** states from the 2016 Particle Data Group Data Book \cite{PDG16}). Deviations of the HRG calculations from the lattice curves of flavour specific susceptibilities at specific temperatures in the crossover region may be affected by the inclusion of certain states \cite{Jaki20141,Stafford,Bazavov2014}, thus a realistic determination of the underlying hadronic spectrum is key to this study. The PDG2016\Plus{} hadronic spectrum has been shown to be an optimized compromise between too few (found) and too many (from a simple Quark Model) excited states when compared to a large number of lattice QCD predictions \cite{PDG2016+}. 

Yield data for $\pi^{\text{+}}$, $ \pi^{\text{-}}$, $K^{\text{+}}$, $K^{\text{-}}$, $p$,  $\bar{p}$, $\Lambda$, $\bar{\Lambda}$, $\Xi^{-}$, $\bar{\Xi}^{+}$, $\Omega^{-}$, $\bar{\Omega}^{+}$, $K_{\rm{S}}^{0},$ and $\phi$ for ALICE PbPb collisions at $\sqrt{s_{\rm{NN}}} = 2.76$ TeV \cite{PbPb276,PbPb276K0S,PbPb276phi,PbPb276MultiS} and preliminary results at $5.02$ TeV \cite{BELLINI2019427} in the 0 - 10\% centrality class, as well as  STAR AuAu collisions at $\sqrt{s_{\rm{NN}}} = 11.5, 19.6, 27.0, 39.0, 62.4$ and $200$ GeV \cite{BES,STAR_StrangenessPRC,STARAuAu,STAR624,STAR200,STAR200Hyperons} were used. We excluded AuAu collisions at $\sqrt{s_{\rm{NN}}} = 7.7$ GeV due to the wider centrality binning of the data, particularly for multi-strange baryons.

For the sake of brevity, we introduced a shorthand notation when naming our fits with (anti)particle species (e.g. $\Omega$ refers to both $\Omega^{-} $ and $\bar{\Omega}^+$, etc.). This shorthand is used for the remainder of this letter.

(Anti)proton yields for the STAR data in Refs. \cite{BES,STAR624,STAR200} are all inclusive. In order to correct for weak-decay feed-down contributions from ($\bar{\Lambda}$)$\Lambda$, we interpolated the contributions to (anti)proton yields based on the method suggested in Ref. \cite{Feedown}. Table \cref{tab:feeddown} summarizes our interpolation results. The contribution from ($\bar{\Lambda}$)$\Lambda$ to (anti)proton yields are labeled as $\delta$. These $\delta$ values were subtracted from unity and multiplied by their respective (anti)proton yields. The resulting (anti)proton yields were then used for the entirety of this analysis. This procedure should be considered an upper limit for the feed-down contribution since the experiment imposes an, albeit loose, primary vertex cut on the ($\bar{\Lambda}$)$\Lambda$ decay daughter candidates. Nevertheless, these percentages are in general agreement with estimates in the aforementioned STAR papers.

\begin{table}[htbp]
\caption{Interpolation results from methods used in Ref. \cite{Feedown} for weak-decay feed-down contributions to (anti)proton yields at STAR AuAu Collisions from $\sqrt{s_{\rm{NN}}} = $ 11.5 to 200 GeV.}

\setlength{\tabcolsep}{9pt}
\centering
\begin{tabular}{@{}ccc@{}}
\toprule
        &  Proton       & Anti-Proton             \\
$\sqrt{s_{_{\rm{NN}}}}$ (GeV)            & $\delta$                                    & $\delta$                     \\ \midrule
\multicolumn{1}{c}{11.5} & \multicolumn{1}{c}{23.00$\%$} &  \multicolumn{1}{c}{48.00$\%$}       \\
\multicolumn{1}{c}{19.6} & \multicolumn{1}{c}{27.50$\%$} &  \multicolumn{1}{c}{44.00$\%$}       \\
\multicolumn{1}{c}{27.0} & \multicolumn{1}{c}{29.50$\%$} &  \multicolumn{1}{c}{41.50$\%$}       \\
\multicolumn{1}{c}{39.0} & \multicolumn{1}{c}{31.00$\%$} &  \multicolumn{1}{c}{40.00$\%$}     \\
\multicolumn{1}{c}{64.2} & \multicolumn{1}{c}{32.00$\%$} &  \multicolumn{1}{c}{38.50$\%$}       \\
\multicolumn{1}{c}{200}  & \multicolumn{1}{c}{34.00$\%$} &  \multicolumn{1}{c}{36.50$\%$}      \\ \bottomrule
\end{tabular}
\label{tab:feeddown}
\end{table}

All our thermal fits were performed with $T_{\rm{ch}}$ (MeV) and V (fm$^{3}$) as free parameters, setting $\gamma_{\mathrm{S}}$ and $\gamma_{\mathrm{q}}$ to unity. Our analysis focused on varying the particle species included in the fit. The particle species included in our temperature fits were $\pi K p$ (light), $\pi K p \Lambda \Xi \Omega K^{0}_{\rm{S}} \phi$ (all) and $K \Lambda \Xi \Omega K^{0}_{\rm{S}} \phi$ (strange), respectively. The inclusion of the kaons in the light fit was done in order to avoid too few degrees of freedom in the fit and has no effect on the extracted freeze-out temperature, since the kaon yield is rather insensitive to the temperature, as was also shown previously in Ref. \cite{Magestro_2002}. For the two ALICE energies, we let $\mu_{\rm{B}} = 0$. For all fits to STAR AuAu data, we let $\mu_{\rm{B}}$ be a free parameter in the fits to also gauge its sensitivity to the flavour-specific fits.

\section{Results and Discussion}
\begin{figure}[htbp]
\centering
\includegraphics[width=.99\linewidth, trim = {0 .3cm 0 0}, clip]{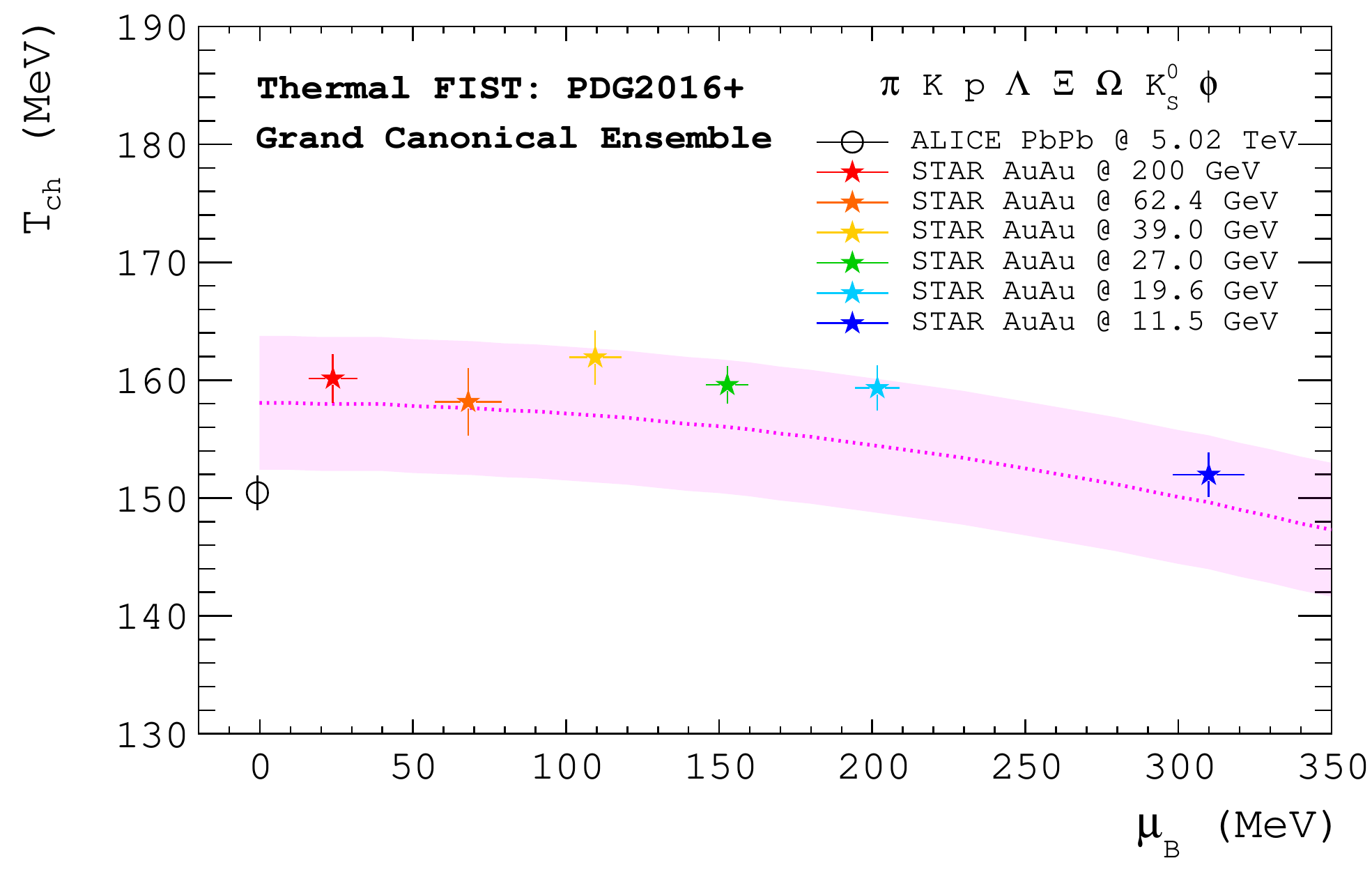}

\caption{\textit{Full} GCE fits to STAR and ALICE data measured at collision energies ranging from $\sqrt{s_{\rm{NN}}} = $ 11.5 GeV to 5.02 TeV (0 - 10\%) using The FIST with the PDG2016\Plus{} hadronic spectrum. Magenta bands shows a spline fit to the points.}
\label{fig:fullfitsmuB}
\end{figure}

We extracted freeze-out parameters, $T_{\mathrm{ch}}$ and $V$, for the full ($\pi K p \Lambda \Xi \Omega K^{0}_{\rm{S}} \phi$), light ($\pi K p$), and strange ($K \Lambda \Xi \Omega K^{0}_{\rm{S}} \phi$) particle thermal fits. Figure \cref{fig:fullfitsmuB} shows $T_{\mathrm{ch}}$ from the full fits as a function of $\mu_{\rm{B}}$. The magenta line represents the spline fit function with three nodes (PbPb at $\sqrt{s_{\rm{NN}}} = 5.02$ TeV and AuAu at $\sqrt{s_{\rm{NN}}} = 27.0$ and $19.6$ GeV) to the freeze-out parameters. The magenta band represents the residual to the spline fit and is determined by the mean square error of the spline function.

\begin{figure}[htbp]
\centering
\includegraphics[width=1.00\linewidth, trim = {0 .5cm 0 0}, clip]{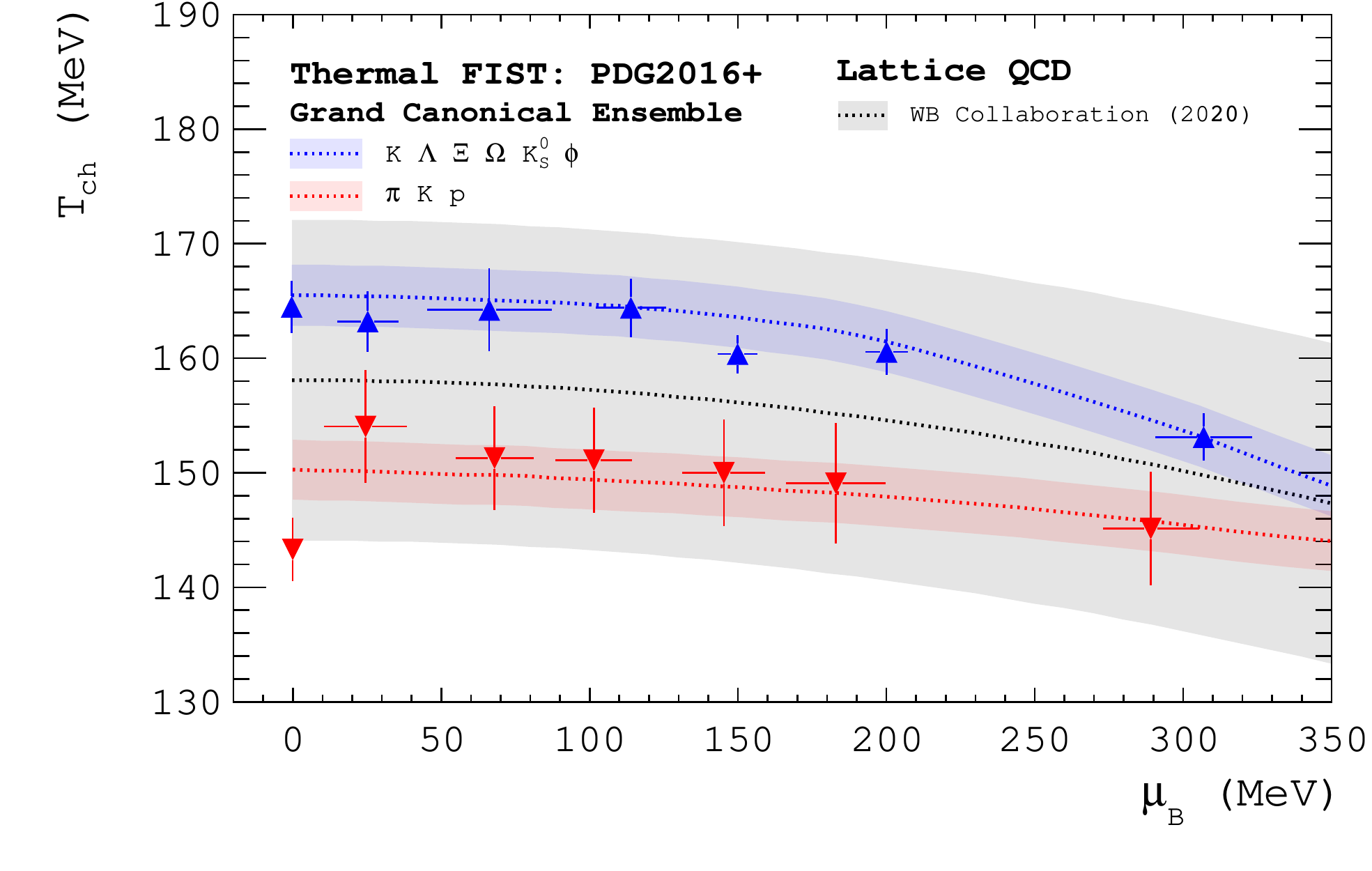}

\caption{\textit{Strange} (blue points) and \textit{light} (red points) GCE fits to STAR and ALICE data measured at collision energies ranging from $\sqrt{s_{\rm{NN}}} = $ 11.5 GeV to 5.02 TeV (0 - 10$\%$) via The FIST using the PDG2016\Plus{} hadronic spectrum.}
\label{fig:fitsvlatticemuB}
\end{figure}

Figure \cref{fig:fitsvlatticemuB} shows $T_{\mathrm{ch}}$ from the light and strange fit as a function of $\mu_{\rm{B}}$; both fits are compared to Lattice QCD calculations in Ref. \cite{WBJuly2020}. Each flavour dependent spline fit and error band was determined in the same manner as in Figure \cref{fig:fullfitsmuB}. The width of the lattice curve is based on the width ($\sigma$) of the chiral susceptibility \cite{WBJuly2020}.

Detailed fit results for each energy including $V$ and $\chi^{2}/dof$, are shown in Table \cref{tab:thermfitmu}. Generally, the separation into light and strange particles improves the quality of the fits by at least a factor two at all energies. 

In Figure \cref{fig:fitsvlatticemuB}, the light and strange fits consistently fall within the lattice QCD crossover width as determined by detailed studies of the temperature dependence of the chiral order parameters \cite{WBJuly2020}. Our flavour-dependent fits agree with the calculated freeze-out temperatures from net-proton, net-charge and net-kaon fluctuations up to $\mu_{\rm{B}}\simeq150$ MeV \cite{AlbaNetQ,NetK}. $T_{\mathrm{ch}}$ remains constant with increasing $\mu_{\rm{B}}$ until $\mu_{\rm{B}}\simeq100$ MeV, where the strange and light fit begin to approach. The two fits converge within errors at $\mu_{\rm{B}}\simeq300$ MeV; therefore, we propose that a separate treatment of strange and light particles might not be meaningful at $\mu_{\mathrm{B}} \geq 300$ MeV. The convergence of the two flavour-dependent temperatures is expected in the vicinity of a critical point in the QCD phase diagram.

\begin{figure}[htbp]
\centering
\includegraphics[width=0.99\linewidth, trim = {0.6cm 1cm 1.9cm 0}, clip]{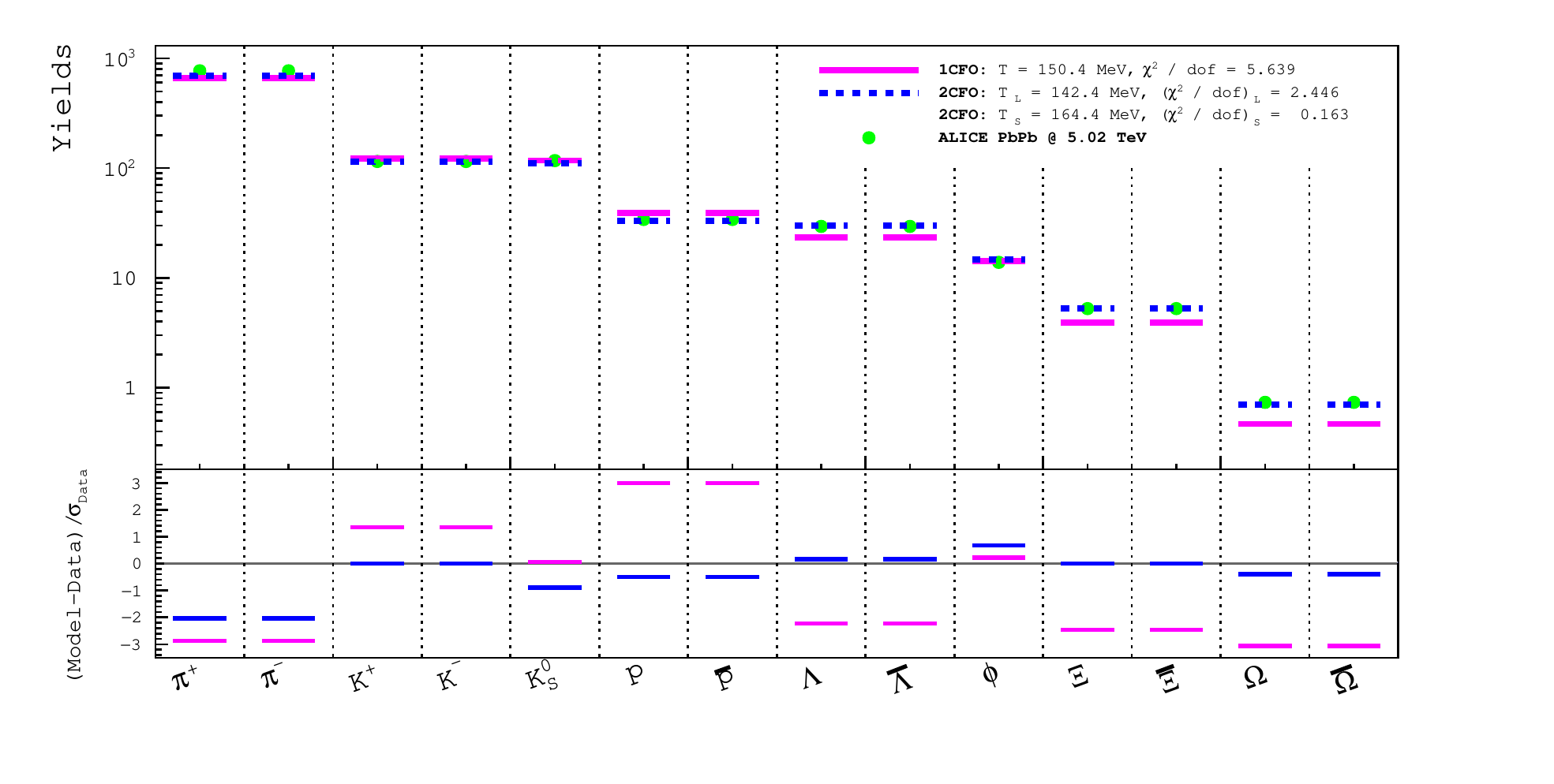}
\includegraphics[width=0.99\linewidth, trim = {.6cm 1cm 1.9cm 0}, clip]{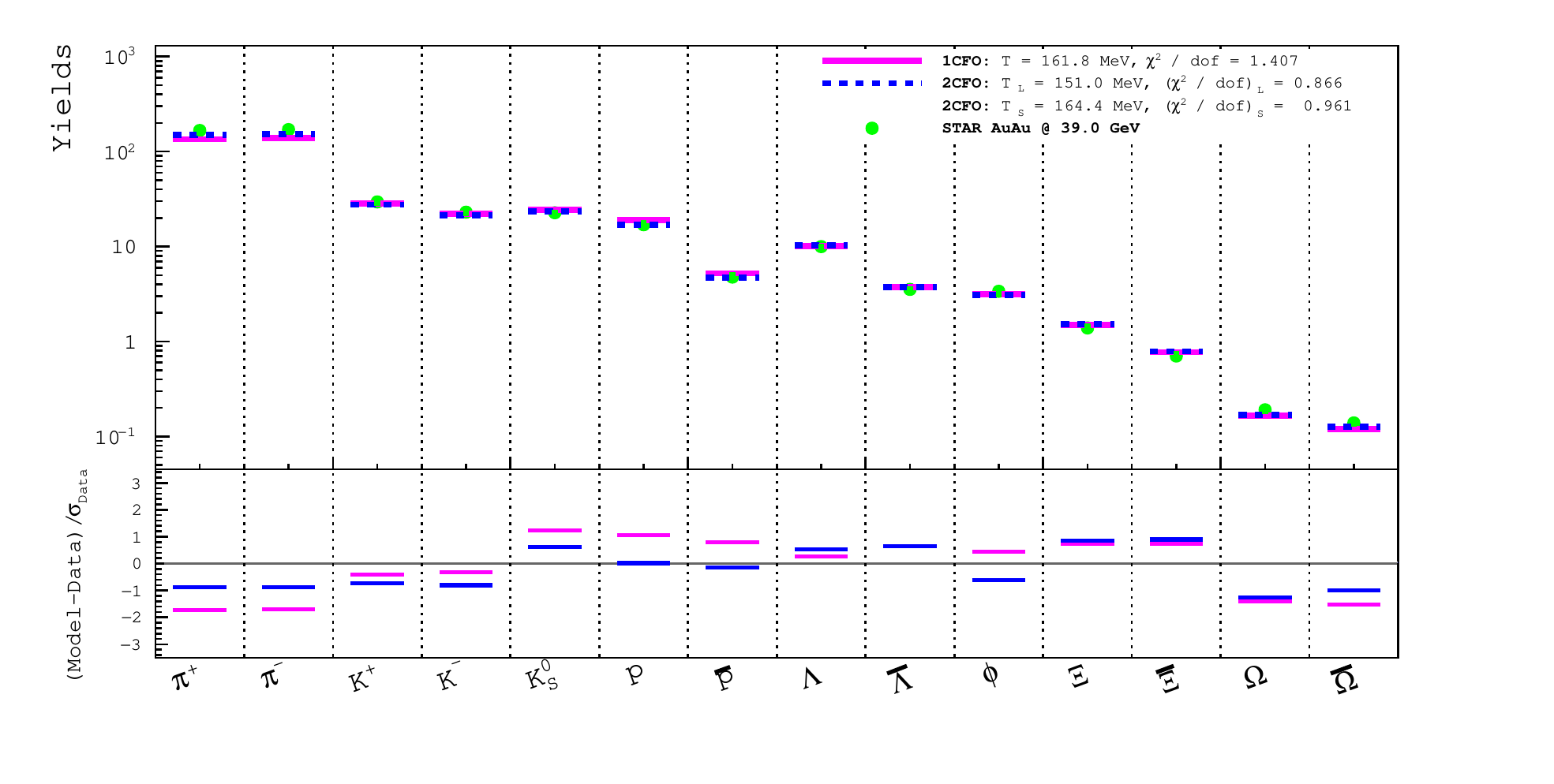}
\caption{Top and bottom panels show GCE fits to ALICE PbPb at $\sqrt{s_{\rm{NN}}} = 5.02$ TeV (0 - 10\%) and STAR AuAu at $\sqrt{s_{\rm{NN}}} = 39.0$ GeV (0 - 10\%), respectively, via The FIST using the PDG2016\Plus{} hadronic spectrum. Single temperature (1CFO) yield calculations are shown in magenta. Two temperature (2CFO) yield calculations are shown in dashed blue lines. Experimental values \cite{BELLINI2019427,STAR_StrangenessPRC} are shown in green.}
\label{fig:yieldfits}
\end{figure}

\begin{table*}[htbp]
\caption{The FIST Grand Canonical Ensemble Yield Fits via the PDG2016\Plus{}  hadronic spectrum for collision energies ranging from $\sqrt{s_{\rm{NN}}} = 11.5$ GeV to $5.02$ TeV. The top, middle and bottom sections show the full ($\pi K p \Lambda \Xi \Omega K^0_S \phi$), light ($\pi K p$), and strange ($K \Lambda \Xi \Omega K^0_S \phi$) particle fits, respectively. For all fits, $\mu_B$,  $T_\mathrm{ch}$, and $V$ were used as free parameters.}
\begin{center}
\setlength{\tabcolsep}{19pt}
\renewcommand{\arraystretch}{1}
\begin{tabular}{@{}ccccc@{}}
\toprule
 & & $\pi K p \Lambda \Xi \Omega K^0_S \phi$ & & \vspace{.5em} \\
 $\sqrt{s_{\rm{NN}}} $ (GeV) & $\mu_{\rm{B}}$ (MeV) & $T_{\mathrm{ch}}$ (MeV)  & $V(fm^3)$  & $\chi^{2}/dof$ \\
\midrule
 $5020$ & 0.0 & 150.4 $\pm$ 1.50  & 6238.8 $\pm$ 538.8 & 78.9/12 \\ 
 $2760$ & 0.0 & 149.6 $\pm$ 1.76  & 5764.4 $\pm$ 635.8 & 23.4/12 \\ 
 $200$ & 25.0 $\pm$ 7.94 & 160.0 $\pm$ 2.07  & 1596.3 $\pm$ 198.3 & 23.3/10 \\ 
 $62.4$ & 68.4 $\pm$ 11.0 & 158.1 $\pm$ 2.87  & 1151.7 $\pm$ 178.2 & 41.0/11 \\ 
 $39$ & 110.3 $\pm$ 8.51 & 161.8 $\pm$ 2.32  & 751.6 $\pm$ 103.0 & 15.5/11 \\ 
 $27$ & 154.2 $\pm$ 6.92 & 159.4 $\pm$ 1.60  & 741.8 $\pm$ 70.4 & 12.8/11 \\ 
 $19.6$ & 202.8 $\pm$ 7.29 & 159.3 $\pm$ 1.91  & 643.2 $\pm$ 72.6 & 15.2/11 \\ 
 $11.5$ & 310.7 $\pm$ 11.7 & 151.8 $\pm$ 1.89  & 649.4 $\pm$ 78.8 & 15.9/11 \\ 
\midrule
& & $\pi K p$ & & \vspace{.5em} \\

 $\sqrt{s_{\rm{NN}}}$ (GeV) & $\mu_{\rm{B}}$ (MeV) &  $T_\mathrm{ch}$ (MeV)  & $V(fm^3)$  & $\chi^{2}/dof$ \\
 \midrule
 $5020$ & 0.0 & 142.4 $\pm$ 1.70  & 9371.6 $\pm$ 902.1 & 14.6/4 \\ 
$2760$ & 0.0 & 143.2 $\pm$ 2.79  & 8031.7 $\pm$ 1263.0 & 5.65/4 \\ 
$200$ &  24.5 $\pm$ 14.0 & 153.9 $\pm$ 4.91  & 2210.0 $\pm$ 553.4 & 3.10/3 \\ 
$62.4$ & 67.8 $\pm$ 13.1 & 151.2 $\pm$ 4.53 & 1721.7 $\pm$ 394.5 & 7.79/3 \\ 
$39$ & 101.4 $\pm$ 12.9 & 151.0 $\pm$ 4.61 & 1368.8 $\pm$ 338.9 & 2.60/3 \\ 
$27$ & 145.3 $\pm$ 14.0 & 149.9 $\pm$ 4.68 & 1333.3 $\pm$ 336.8 & 2.79/3 \\ 
$19.6$ & 182.9 $\pm$ 16.7 & 149.0 $\pm$ 5.24 & 1186.1 $\pm$ 341.5 & 7.66/3 \\ 
$11.5$ & 289.4 $\pm$ 16.2 & 145.0 $\pm$ 4.98 & 1074.8 $\pm$ 300.1 & 2.67/3 \\ 
\midrule
& & $ K \Lambda \Xi \Omega K^0_S \phi$ & & \vspace{.5em} \\
 $\sqrt{s_{\rm{NN}}}$ (GeV) & $\mu_\mathrm{B}$ (MeV) & $T_\mathrm{ch}$ (MeV) & $V(fm^3)$  & $\chi^{2}/dof$ \\
 \midrule
 $5020$ & 0.0 & 164.4 $\pm$ 2.28 & 3086.8 $\pm$ 366.7 & 1.58/8 \\ 
$2760$ & 0.0 & 153.9 $\pm$ 2.30 & 4389.7 $\pm$ 640.8 & 10.5/8 \\ 
$200$ & 25.7 $\pm$ 10.3  & 163.2 $\pm$ 2.64  & 1287.5 $\pm$ 208.8 & 17.3/6 \\ 
$62.4$ & 66.7 $\pm$ 21.0  & 164.2 $\pm$ 3.63 & 784.5 $\pm$ 152.2 & 21.4/7 \\ 
$39$ & 116.0 $\pm$ 11.7 & 164.4 $\pm$ 2.57  & 643.6 $\pm$ 97.7 & 6.73/7 \\ 
$27$ & 156.0 $\pm$ 7.88 & 160.4 $\pm$ 1.68  & 695.0 $\pm$ 69.3 & 3.55/7 \\ 
$19.6$ & 206.5 $\pm$ 8.30 & 160.7 $\pm$ 2.00  & 585.2 $\pm$ 69.3 & 3.87/7 \\ 
$11.5$ & 321.0 $\pm$ 14.7 & 153.4 $\pm$ 2.08 & 574.8 $\pm$ 77.2  & 5.36/7 \\ 
 \bottomrule
\end{tabular}
\end{center}
\label{tab:thermfitmu}
\end{table*}





The greatest difference between the two temperatures is seen at the highest energy, namely the ALICE top energy. To address the impact of a flavour dependent freeze-out on the particle abundances, we calculated yields for the full particle set ($\pi K p \Lambda \Xi \Omega K^0_{\rm{S}} \phi$) for ALICE PbPb 5.02 TeV with a one chemical freeze-out (1CFO) approach and compared them with yields calculated for light ($\pi K p$) and strange ($ K \Lambda \Xi \Omega K^0_{\rm{S}} \phi$) particles separately with a two chemical freeze-out (2CFO) approach.  We fixed the temperature(s) and volume(s) to the $T_{\rm{ch}}$ and $V$ values shown in Table \cref{tab:thermfitmu} for $\sqrt{s_{\rm{NN}}} = 5020$ GeV. In the 1CFO approach, we calculated yields using a temperature of 150.4 MeV.  In the 2CFO approach, our light and strange particle yield calculations were done with temperatures fixed to 142.4 MeV and 164.4 MeV, respectively. We note that our 1CFO temperature differs from the value quoted by ALICE, which is based on the Heidelberg-GSI model, by about 4 MeV, most likely due to the difference in the hadronic input spectrum \cite{BELLINI2019427}.

Figure \cref{fig:yieldfits} shows a comparison between the 1CFO and 2CFO approaches for data sets at two exemplary energies, namely the preliminary 5.02 TeV central PbPb data from ALICE and the 39.0 GeV central AuAu data from STAR. The deviations of each yield calculation from the experimental value are shown at the bottom of each plot. We observe that the 2CFO approach provides an excellent and much improved description of the experimental data;  rendering yields within one standard deviation of the experimental measurements for most particle species. The 2CFO treatment all but eliminates the tension between light and strange baryons, the so-called proton anomaly, seen in the 1CFO approach. It should also be noted that alternative methods to treat interactions in the SHM via the S-matrix approach \cite{Andronic:2018qqt} impact in particular the proton yields and improve the performance of the 1CFO method in the Heidelberg-GSI fits \cite{PBMNature}.

\begin{figure}[htbp]
\centering
\includegraphics[width=1.00\linewidth]{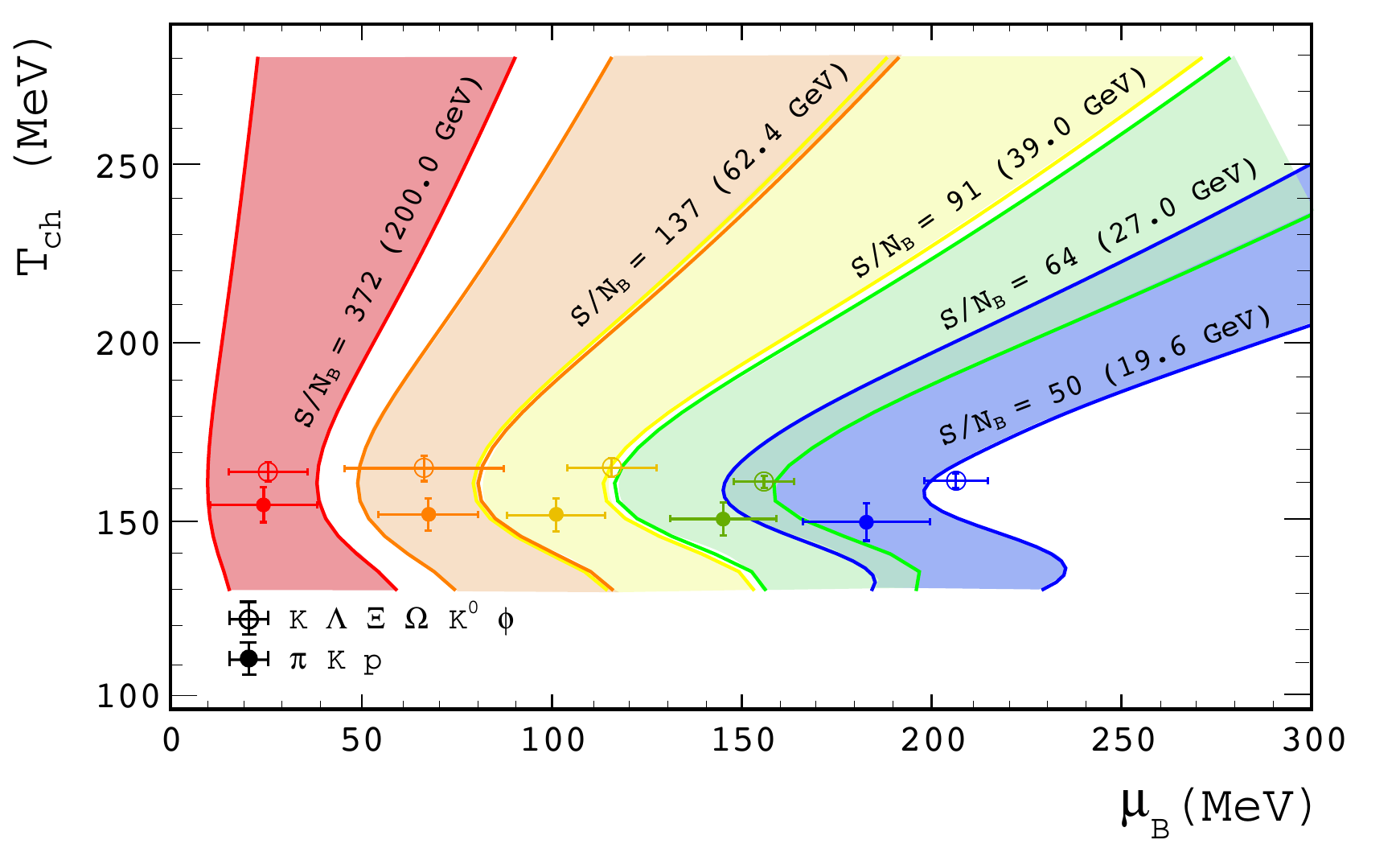}

\caption{Isentropic trajectories (fixed S/N$_{B}$) in the T-$\mu_{B}$ plane based on a lattice QCD Equation of State expanded to finite density \cite{Gunther:2016vcp}. The open points show the \textit{light} hadron freeze-out parameters and the closed points show the \textit{strange} hadron freeze-out parameters in our 2CFO approach. The S/N$_{B}$ ratios correspond to the RHIC energies 200, 62.4, 39, 27, 19.6 GeV, respectively.}
\label{fig:STAR_Isentropes}
\end{figure}

As a cross-check with lattice QCD predictions, we determined whether our 2CFO parameters lie on the isentropic trajectories in the T-$\mu_{b}$ plane, which were calculated using a lattice QCD equation of state \cite{Gunther:2016vcp}. The validity of this approach to finite densities has been proven out to $\mu_{b}$/T = 2. Therefore, in Figure \cref{fig:STAR_Isentropes}, we show our data only for collision energies down to $\sqrt(s_{NN})$ = 19.6 GeV. The ALICE points are also excluded since at zero baryo-chemical potential the S/N$_{B}$ value diverges. For all five collision energies light and strange freeze-out parameters lie well within the calculated trajectories. The uncertainties on the isentropic trajectories are based on folding the errors of the light hadron freeze-out parameters for $T_\mathrm{ch}$ and $\mu_{\rm{B}}$ into the calculation. A trend towards a sharper turn in the isentropic curve seems to develop with increasing density. It is likely that in order to capture said turn in future lattice QCD calculations beyond $\mu_{\rm{B}}$/T = 2, it will be necessary to extend the Taylor expansion to terms higher than c$_{6}$ \cite{Gunther:2016vcp}.

\section{Conclusion}
We presented calculations of the chemical freeze-out temperature ($T_\mathrm{ch}$) based on particle yields from STAR and ALICE measured at collision energies ranging from $\sqrt{s_{\rm{NN}}} = $ 11.5 GeV to 5.02 TeV. Based on a splined fit to our thermal fit parameters at all energies, we calculated a light flavour freeze-out temperature $T_\mathrm{L} = 150.2 \pm 2.6$ MeV and a strange flavour freeze-out temperature $T_\mathrm{S} = 165.1 \pm 2.7$ MeV at vanishing $\mu_{\rm{B}}$, employing the GCE approach within the framework of the The FIST HRG model package. We showed evidence for flavour-dependent chemical freeze-out temperatures in the crossover region of the QCD phase diagram, which start to converge above $\mu_{\rm{B}} \simeq300$ MeV. We employed the flavour-dependent two temperature approach via The FIST to successfully model and reproduce experimental yields at top ALICE energies. Thus, at the highest energies at RHIC and the LHC a separation of the hadronization temperature of light and strange particles seems likely. Furthermore, our results suggest the existence of a critical point in the QCD phase diagram above $\mu_{\rm{B}} \simeq300$ MeV and a temperature below $T_\mathrm{ch}\simeq150$ MeV.

\section{Acknowledgments}
The authors acknowledge edifying discussions with Volodymyr Vovchenko, Claudia Ratti, Paolo Parotto, Livio Bianchi, Boris Hippolyte and Jamie Stafford -- particularly for her assistance in producing the lattice QCD based isentropes in Figure \cref{fig:STAR_Isentropes}. This work was supported by the DOE grant DEFG02-07ER4152.

\bibliographystyle{apsrev4-1}
\bibliography{refs}

\begin{thebibliography}{38}%
\makeatletter
\providecommand \@ifxundefined [1]{%
 \@ifx{#1\undefined}
}%
\providecommand \@ifnum [1]{%
 \ifnum #1\expandafter \@firstoftwo
 \else \expandafter \@secondoftwo
 \fi
}%
\providecommand \@ifx [1]{%
 \ifx #1\expandafter \@firstoftwo
 \else \expandafter \@secondoftwo
 \fi
}%
\providecommand \natexlab [1]{#1}%
\providecommand \enquote  [1]{``#1''}%
\providecommand \bibnamefont  [1]{#1}%
\providecommand \bibfnamefont [1]{#1}%
\providecommand \citenamefont [1]{#1}%
\providecommand \href@noop [0]{\@secondoftwo}%
\providecommand \href [0]{\begingroup \@sanitize@url \@href}%
\providecommand \@href[1]{\@@startlink{#1}\@@href}%
\providecommand \@@href[1]{\endgroup#1\@@endlink}%
\providecommand \@sanitize@url [0]{\catcode `\\12\catcode `\$12\catcode
  `\&12\catcode `\#12\catcode `\^12\catcode `\_12\catcode `\%12\relax}%
\providecommand \@@startlink[1]{}%
\providecommand \@@endlink[0]{}%
\providecommand \url  [0]{\begingroup\@sanitize@url \@url }%
\providecommand \@url [1]{\endgroup\@href {#1}{\urlprefix }}%
\providecommand \urlprefix  [0]{URL }%
\providecommand \Eprint [0]{\href }%
\providecommand \doibase [0]{http://dx.doi.org/}%
\providecommand \selectlanguage [0]{\@gobble}%
\providecommand \bibinfo  [0]{\@secondoftwo}%
\providecommand \bibfield  [0]{\@secondoftwo}%
\providecommand \translation [1]{[#1]}%
\providecommand \BibitemOpen [0]{}%
\providecommand \bibitemStop [0]{}%
\providecommand \bibitemNoStop [0]{.\EOS\space}%
\providecommand \EOS [0]{\spacefactor3000\relax}%
\providecommand \BibitemShut  [1]{\csname bibitem#1\endcsname}%
\let\auto@bib@innerbib\@empty
\bibitem [{\citenamefont {Bellwied}\ \emph {et~al.}(2015)\citenamefont
  {Bellwied}, \citenamefont {Bors{\'a}nyi}, \citenamefont {Fodor},
  \citenamefont {G{\"u}nther}, \citenamefont {Katz}, \citenamefont {Ratti},\
  and\ \citenamefont {Szab{\'o}}}]{Lattice}%
  \BibitemOpen
  \bibfield  {author} {\bibinfo {author} {\bibfnamefont {R.}~\bibnamefont
  {Bellwied}}, \bibinfo {author} {\bibfnamefont {S.}~\bibnamefont
  {Bors{\'a}nyi}}, \bibinfo {author} {\bibfnamefont {Z.}~\bibnamefont {Fodor}},
  \bibinfo {author} {\bibfnamefont {J.}~\bibnamefont {G{\"u}nther}}, \bibinfo
  {author} {\bibfnamefont {S.}~\bibnamefont {Katz}}, \bibinfo {author}
  {\bibfnamefont {C.}~\bibnamefont {Ratti}}, \ and\ \bibinfo {author}
  {\bibfnamefont {K.}~\bibnamefont {Szab{\'o}}},\ }\href {\doibase
  https://doi.org/10.1016/j.physletb.2015.11.011} {\bibfield  {journal}
  {\bibinfo  {journal} {Physics Letters B}\ }\textbf {\bibinfo {volume}
  {751}},\ \bibinfo {pages} {559 } (\bibinfo {year} {2015})}\BibitemShut
  {NoStop}%
\bibitem [{\citenamefont {Bors{\'a}nyi}\ \emph {et~al.}(2010)\citenamefont
  {Bors{\'a}nyi}, \citenamefont {Fodor}, \citenamefont {Hoelbling},
  \citenamefont {Katz}, \citenamefont {Krieg}, \citenamefont {Ratti},\ and\
  \citenamefont {Szab{\'o}}}]{WB1}%
  \BibitemOpen
  \bibfield  {author} {\bibinfo {author} {\bibfnamefont {S.}~\bibnamefont
  {Bors{\'a}nyi}}, \bibinfo {author} {\bibfnamefont {Z.}~\bibnamefont {Fodor}},
  \bibinfo {author} {\bibfnamefont {C.}~\bibnamefont {Hoelbling}}, \bibinfo
  {author} {\bibfnamefont {S.~D.}\ \bibnamefont {Katz}}, \bibinfo {author}
  {\bibfnamefont {S.}~\bibnamefont {Krieg}}, \bibinfo {author} {\bibfnamefont
  {C.}~\bibnamefont {Ratti}}, \ and\ \bibinfo {author} {\bibfnamefont {K.~K.}\
  \bibnamefont {Szab{\'o}}},\ }\href {\doibase 10.1007/JHEP09(2010)073}
  {\bibfield  {journal} {\bibinfo  {journal} {Journal of High Energy Physics}\
  }\textbf {\bibinfo {volume} {2010}},\ \bibinfo {pages} {73} (\bibinfo {year}
  {2010})}\BibitemShut {NoStop}%
\bibitem [{\citenamefont {Bors{\'a}nyi}\ \emph {et~al.}(2014)\citenamefont
  {Bors{\'a}nyi}, \citenamefont {Fodor}, \citenamefont {Hoelbling},
  \citenamefont {Katz}, \citenamefont {Krieg},\ and\ \citenamefont
  {Szab{\'o}}}]{WB2}%
  \BibitemOpen
  \bibfield  {author} {\bibinfo {author} {\bibfnamefont {S.}~\bibnamefont
  {Bors{\'a}nyi}}, \bibinfo {author} {\bibfnamefont {Z.}~\bibnamefont {Fodor}},
  \bibinfo {author} {\bibfnamefont {C.}~\bibnamefont {Hoelbling}}, \bibinfo
  {author} {\bibfnamefont {S.~D.}\ \bibnamefont {Katz}}, \bibinfo {author}
  {\bibfnamefont {S.}~\bibnamefont {Krieg}}, \ and\ \bibinfo {author}
  {\bibfnamefont {K.~K.}\ \bibnamefont {Szab{\'o}}},\ }\href {\doibase
  https://doi.org/10.1016/j.physletb.2014.01.007} {\bibfield  {journal}
  {\bibinfo  {journal} {Physics Letters B}\ }\textbf {\bibinfo {volume}
  {730}},\ \bibinfo {pages} {99 } (\bibinfo {year} {2014})}\BibitemShut
  {NoStop}%
\bibitem [{\citenamefont {Bazavov}\ \emph
  {et~al.}(2014{\natexlab{a}})\citenamefont {Bazavov} \emph
  {et~al.}}]{HotQCD1}%
  \BibitemOpen
  \bibfield  {author} {\bibinfo {author} {\bibfnamefont {A.}~\bibnamefont
  {Bazavov}} \emph {et~al.} (\bibinfo {collaboration} {HotQCD Collaboration}),\
  }\href {\doibase 10.1103/PhysRevD.90.094503} {\bibfield  {journal} {\bibinfo
  {journal} {Phys. Rev. D}\ }\textbf {\bibinfo {volume} {90}},\ \bibinfo
  {pages} {094503} (\bibinfo {year} {2014}{\natexlab{a}})}\BibitemShut
  {NoStop}%
\bibitem [{\citenamefont {Stachel}\ \emph {et~al.}(2014)\citenamefont
  {Stachel}, \citenamefont {Andronic}, \citenamefont {Braun-Munzinger},\ and\
  \citenamefont {Redlich}}]{Stachel_2014}%
  \BibitemOpen
  \bibfield  {author} {\bibinfo {author} {\bibfnamefont {J.}~\bibnamefont
  {Stachel}}, \bibinfo {author} {\bibfnamefont {A.}~\bibnamefont {Andronic}},
  \bibinfo {author} {\bibfnamefont {P.}~\bibnamefont {Braun-Munzinger}}, \ and\
  \bibinfo {author} {\bibfnamefont {K.}~\bibnamefont {Redlich}},\ }\href
  {\doibase 10.1088/1742-6596/509/1/012019} {\bibfield  {journal} {\bibinfo
  {journal} {Journal of Physics: Conference Series}\ }\textbf {\bibinfo
  {volume} {509}},\ \bibinfo {pages} {012019} (\bibinfo {year}
  {2014})}\BibitemShut {NoStop}%
\bibitem [{\citenamefont {Cleymans}\ and\ \citenamefont
  {Redlich}(1998)}]{Cleymans1998}%
  \BibitemOpen
  \bibfield  {author} {\bibinfo {author} {\bibfnamefont {J.}~\bibnamefont
  {Cleymans}}\ and\ \bibinfo {author} {\bibfnamefont {K.}~\bibnamefont
  {Redlich}},\ }\href {\doibase 10.1103/PhysRevLett.81.5284} {\bibfield
  {journal} {\bibinfo  {journal} {Phys. Rev. Lett.}\ }\textbf {\bibinfo
  {volume} {81}},\ \bibinfo {pages} {5284} (\bibinfo {year}
  {1998})}\BibitemShut {NoStop}%
\bibitem [{\citenamefont {Andronic}\ \emph {et~al.}(2018)\citenamefont
  {Andronic}, \citenamefont {Braun-Munzinger}, \citenamefont {Redlich},\ and\
  \citenamefont {Stachel}}]{PBMNature}%
  \BibitemOpen
  \bibfield  {author} {\bibinfo {author} {\bibfnamefont {A.}~\bibnamefont
  {Andronic}}, \bibinfo {author} {\bibfnamefont {P.}~\bibnamefont
  {Braun-Munzinger}}, \bibinfo {author} {\bibfnamefont {K.}~\bibnamefont
  {Redlich}}, \ and\ \bibinfo {author} {\bibfnamefont {J.}~\bibnamefont
  {Stachel}},\ }\href {\doibase https://doi.org/10.1038/s41586-018-0491-6}
  {\bibfield  {journal} {\bibinfo  {journal} {Nature}\ }\textbf {\bibinfo
  {volume} {561}},\ \bibinfo {pages} {321 } (\bibinfo {year}
  {2018})}\BibitemShut {NoStop}%
\bibitem [{\citenamefont {Ratti}\ \emph {et~al.}(2012)\citenamefont {Ratti},
  \citenamefont {Bellwied}, \citenamefont {Cristoforetti},\ and\ \citenamefont
  {Barbaro}}]{Ratti2012}%
  \BibitemOpen
  \bibfield  {author} {\bibinfo {author} {\bibfnamefont {C.}~\bibnamefont
  {Ratti}}, \bibinfo {author} {\bibfnamefont {R.}~\bibnamefont {Bellwied}},
  \bibinfo {author} {\bibfnamefont {M.}~\bibnamefont {Cristoforetti}}, \ and\
  \bibinfo {author} {\bibfnamefont {M.}~\bibnamefont {Barbaro}},\ }\href
  {\doibase 10.1103/PhysRevD.85.014004} {\bibfield  {journal} {\bibinfo
  {journal} {Phys. Rev. D}\ }\textbf {\bibinfo {volume} {85}},\ \bibinfo
  {pages} {014004} (\bibinfo {year} {2012})}\BibitemShut {NoStop}%
\bibitem [{\citenamefont {Bellwied}\ \emph {et~al.}(2013)\citenamefont
  {Bellwied}, \citenamefont {Borsanyi}, \citenamefont {Fodor}, \citenamefont
  {Katz},\ and\ \citenamefont {Ratti}}]{SuscRat}%
  \BibitemOpen
  \bibfield  {author} {\bibinfo {author} {\bibfnamefont {R.}~\bibnamefont
  {Bellwied}}, \bibinfo {author} {\bibfnamefont {S.}~\bibnamefont {Borsanyi}},
  \bibinfo {author} {\bibfnamefont {Z.}~\bibnamefont {Fodor}}, \bibinfo
  {author} {\bibfnamefont {S.~D.}\ \bibnamefont {Katz}}, \ and\ \bibinfo
  {author} {\bibfnamefont {C.}~\bibnamefont {Ratti}},\ }\href {\doibase
  10.1103/PhysRevLett.111.202302} {\bibfield  {journal} {\bibinfo  {journal}
  {Phys. Rev. Lett.}\ }\textbf {\bibinfo {volume} {111}},\ \bibinfo {pages}
  {202302} (\bibinfo {year} {2013})}\BibitemShut {NoStop}%
\bibitem [{\citenamefont {Karsch}(2012)}]{Karsch2012}%
  \BibitemOpen
  \bibfield  {author} {\bibinfo {author} {\bibfnamefont {F.}~\bibnamefont
  {Karsch}},\ }\href {\doibase 10.2478/s11534-012-0074-3} {\bibfield  {journal}
  {\bibinfo  {journal} {Central European Journal of Physics}\ }\textbf
  {\bibinfo {volume} {10}},\ \bibinfo {pages} {1234} (\bibinfo {year}
  {2012})}\BibitemShut {NoStop}%
\bibitem [{\citenamefont {Adamczyk}\ \emph {et~al.}(2014)\citenamefont
  {Adamczyk} \emph {et~al.}}]{Adamczyk:2013dal}%
  \BibitemOpen
  \bibfield  {author} {\bibinfo {author} {\bibfnamefont {L.}~\bibnamefont
  {Adamczyk}} \emph {et~al.} (\bibinfo {collaboration} {STAR Collaboration}),\
  }\href {\doibase 10.1103/PhysRevLett.112.032302} {\bibfield  {journal}
  {\bibinfo  {journal} {Phys. Rev. Lett.}\ }\textbf {\bibinfo {volume} {112}},\
  \bibinfo {pages} {032302} (\bibinfo {year} {2014})}\BibitemShut {NoStop}%
\bibitem [{\citenamefont {Bellwied}\ \emph {et~al.}(2019)\citenamefont
  {Bellwied}, \citenamefont {Noronha-Hostler}, \citenamefont {Parotto},
  \citenamefont {Portillo~Vazquez}, \citenamefont {Ratti},\ and\ \citenamefont
  {Stafford}}]{NetK}%
  \BibitemOpen
  \bibfield  {author} {\bibinfo {author} {\bibfnamefont {R.}~\bibnamefont
  {Bellwied}}, \bibinfo {author} {\bibfnamefont {J.}~\bibnamefont
  {Noronha-Hostler}}, \bibinfo {author} {\bibfnamefont {P.}~\bibnamefont
  {Parotto}}, \bibinfo {author} {\bibfnamefont {I.}~\bibnamefont
  {Portillo~Vazquez}}, \bibinfo {author} {\bibfnamefont {C.}~\bibnamefont
  {Ratti}}, \ and\ \bibinfo {author} {\bibfnamefont {J.~M.}\ \bibnamefont
  {Stafford}},\ }\href {\doibase 10.1103/PhysRevC.99.034912} {\bibfield
  {journal} {\bibinfo  {journal} {Phys. Rev. C}\ }\textbf {\bibinfo {volume}
  {99}},\ \bibinfo {pages} {034912} (\bibinfo {year} {2019})}\BibitemShut
  {NoStop}%
\bibitem [{\citenamefont {Bluhm}\ and\ \citenamefont
  {Nahrgang}(2019)}]{BluhmKaon}%
  \BibitemOpen
  \bibfield  {author} {\bibinfo {author} {\bibfnamefont {M.}~\bibnamefont
  {Bluhm}}\ and\ \bibinfo {author} {\bibfnamefont {M.}~\bibnamefont
  {Nahrgang}},\ }\href {\doibase
  https://doi.org/10.1140/epjc/s10052-019-6661-3} {\bibfield  {journal}
  {\bibinfo  {journal} {Eur. Phys. J. C.}\ }\textbf {\bibinfo {volume} {79}},\
  \bibinfo {pages} {155} (\bibinfo {year} {2019})}\BibitemShut {NoStop}%
\bibitem [{\citenamefont {Adamczyk}\ \emph {et~al.}(2017)\citenamefont
  {Adamczyk} \emph {et~al.}}]{BES}%
  \BibitemOpen
  \bibfield  {author} {\bibinfo {author} {\bibfnamefont {L.}~\bibnamefont
  {Adamczyk}} \emph {et~al.} (\bibinfo {collaboration} {STAR Collaboration}),\
  }\href {\doibase 10.1103/PhysRevC.96.044904} {\bibfield  {journal} {\bibinfo
  {journal} {Phys. Rev. C}\ }\textbf {\bibinfo {volume} {96}},\ \bibinfo
  {pages} {044904} (\bibinfo {year} {2017})}\BibitemShut {NoStop}%
\bibitem [{\citenamefont {Chatterjee}\ \emph {et~al.}(2017)\citenamefont
  {Chatterjee}, \citenamefont {Dash},\ and\ \citenamefont
  {Mohanty}}]{Chatterjee_2017}%
  \BibitemOpen
  \bibfield  {author} {\bibinfo {author} {\bibfnamefont {S.}~\bibnamefont
  {Chatterjee}}, \bibinfo {author} {\bibfnamefont {A.~K.}\ \bibnamefont
  {Dash}}, \ and\ \bibinfo {author} {\bibfnamefont {B.}~\bibnamefont
  {Mohanty}},\ }\href {\doibase 10.1088/1361-6471/aa8857} {\bibfield  {journal}
  {\bibinfo  {journal} {Journal of Physics G: Nuclear and Particle Physics}\
  }\textbf {\bibinfo {volume} {44}},\ \bibinfo {pages} {105106} (\bibinfo
  {year} {2017})}\BibitemShut {NoStop}%
\bibitem [{\citenamefont {Bellwied}\ \emph {et~al.}(2020)\citenamefont
  {Bellwied}, \citenamefont {Bors\'anyi}, \citenamefont {Fodor}, \citenamefont
  {Guenther}, \citenamefont {Noronha-Hostler}, \citenamefont {Parotto},
  \citenamefont {P\'asztor}, \citenamefont {Ratti},\ and\ \citenamefont
  {Stafford}}]{Bellwied_PRD_2020}%
  \BibitemOpen
  \bibfield  {author} {\bibinfo {author} {\bibfnamefont {R.}~\bibnamefont
  {Bellwied}}, \bibinfo {author} {\bibfnamefont {S.}~\bibnamefont
  {Bors\'anyi}}, \bibinfo {author} {\bibfnamefont {Z.}~\bibnamefont {Fodor}},
  \bibinfo {author} {\bibfnamefont {J.~N.}\ \bibnamefont {Guenther}}, \bibinfo
  {author} {\bibfnamefont {J.}~\bibnamefont {Noronha-Hostler}}, \bibinfo
  {author} {\bibfnamefont {P.}~\bibnamefont {Parotto}}, \bibinfo {author}
  {\bibfnamefont {A.}~\bibnamefont {P\'asztor}}, \bibinfo {author}
  {\bibfnamefont {C.}~\bibnamefont {Ratti}}, \ and\ \bibinfo {author}
  {\bibfnamefont {J.~M.}\ \bibnamefont {Stafford}},\ }\href {\doibase
  10.1103/PhysRevD.101.034506} {\bibfield  {journal} {\bibinfo  {journal}
  {Phys. Rev. D}\ }\textbf {\bibinfo {volume} {101}},\ \bibinfo {pages}
  {034506} (\bibinfo {year} {2020})}\BibitemShut {NoStop}%
\bibitem [{\citenamefont {Vovchenko}\ and\ \citenamefont
  {Stoecker}(2019)}]{FIST}%
  \BibitemOpen
  \bibfield  {author} {\bibinfo {author} {\bibfnamefont {V.}~\bibnamefont
  {Vovchenko}}\ and\ \bibinfo {author} {\bibfnamefont {H.}~\bibnamefont
  {Stoecker}},\ }\href {\doibase https://doi.org/10.1016/j.cpc.2019.06.024}
  {\bibfield  {journal} {\bibinfo  {journal} {Computer Physics Communications}\
  }\textbf {\bibinfo {volume} {244}},\ \bibinfo {pages} {295 } (\bibinfo {year}
  {2019})}\BibitemShut {NoStop}%
\bibitem [{\citenamefont {Alba}\ \emph {et~al.}(2017)\citenamefont {Alba},
  \citenamefont {Bellwied}, \citenamefont {Bors\'anyi}, \citenamefont {Fodor},
  \citenamefont {G\"unther}, \citenamefont {Katz}, \citenamefont
  {Mantovani~Sarti}, \citenamefont {Noronha-Hostler}, \citenamefont {Parotto},
  \citenamefont {Pasztor}, \citenamefont {Vazquez},\ and\ \citenamefont
  {Ratti}}]{PDG2016+}%
  \BibitemOpen
  \bibfield  {author} {\bibinfo {author} {\bibfnamefont {P.}~\bibnamefont
  {Alba}}, \bibinfo {author} {\bibfnamefont {R.}~\bibnamefont {Bellwied}},
  \bibinfo {author} {\bibfnamefont {S.}~\bibnamefont {Bors\'anyi}}, \bibinfo
  {author} {\bibfnamefont {Z.}~\bibnamefont {Fodor}}, \bibinfo {author}
  {\bibfnamefont {J.}~\bibnamefont {G\"unther}}, \bibinfo {author}
  {\bibfnamefont {S.~D.}\ \bibnamefont {Katz}}, \bibinfo {author}
  {\bibfnamefont {V.}~\bibnamefont {Mantovani~Sarti}}, \bibinfo {author}
  {\bibfnamefont {J.}~\bibnamefont {Noronha-Hostler}}, \bibinfo {author}
  {\bibfnamefont {P.}~\bibnamefont {Parotto}}, \bibinfo {author} {\bibfnamefont
  {A.}~\bibnamefont {Pasztor}}, \bibinfo {author} {\bibfnamefont {I.~P.}\
  \bibnamefont {Vazquez}}, \ and\ \bibinfo {author} {\bibfnamefont
  {C.}~\bibnamefont {Ratti}},\ }\href {\doibase 10.1103/PhysRevD.96.034517}
  {\bibfield  {journal} {\bibinfo  {journal} {Phys. Rev. D}\ }\textbf {\bibinfo
  {volume} {96}},\ \bibinfo {pages} {034517} (\bibinfo {year}
  {2017})}\BibitemShut {NoStop}%
\bibitem [{\citenamefont {Patrignani}\ \emph {et~al.}(2016)\citenamefont
  {Patrignani} \emph {et~al.}}]{PDG16}%
  \BibitemOpen
  \bibfield  {author} {\bibinfo {author} {\bibfnamefont {C.}~\bibnamefont
  {Patrignani}} \emph {et~al.} (\bibinfo {collaboration} {Particle Data
  Group}),\ }\href {\doibase 10.1088/1674-1137/40/10/100001} {\bibfield
  {journal} {\bibinfo  {journal} {Chin. Phys.}\ }\textbf {\bibinfo {volume}
  {C40}},\ \bibinfo {pages} {100001} (\bibinfo {year} {2016})}\BibitemShut
  {NoStop}%
\bibitem [{\citenamefont {Noronha-Hostler}\ and\ \citenamefont
  {Greiner}(2014)}]{Jaki20141}%
  \BibitemOpen
  \bibfield  {author} {\bibinfo {author} {\bibfnamefont {J.}~\bibnamefont
  {Noronha-Hostler}}\ and\ \bibinfo {author} {\bibfnamefont {C.}~\bibnamefont
  {Greiner}},\ }\href {\doibase
  https://doi.org/10.1016/j.nuclphysa.2014.08.101} {\bibfield  {journal}
  {\bibinfo  {journal} {Nuclear Physics A}\ }\textbf {\bibinfo {volume}
  {931}},\ \bibinfo {pages} {1108 } (\bibinfo {year} {2014})}\BibitemShut
  {NoStop}%
\bibitem [{\citenamefont {Alba}\ \emph {et~al.}(2020)\citenamefont {Alba},
  \citenamefont {Sarti}, \citenamefont {Noronha-Hostler}, \citenamefont
  {Parotto}, \citenamefont {Portillo-Vazquez}, \citenamefont {Ratti},\ and\
  \citenamefont {Stafford}}]{Stafford}%
  \BibitemOpen
  \bibfield  {author} {\bibinfo {author} {\bibfnamefont {P.}~\bibnamefont
  {Alba}}, \bibinfo {author} {\bibfnamefont {V.~M.}\ \bibnamefont {Sarti}},
  \bibinfo {author} {\bibfnamefont {J.}~\bibnamefont {Noronha-Hostler}},
  \bibinfo {author} {\bibfnamefont {P.}~\bibnamefont {Parotto}}, \bibinfo
  {author} {\bibfnamefont {I.}~\bibnamefont {Portillo-Vazquez}}, \bibinfo
  {author} {\bibfnamefont {C.}~\bibnamefont {Ratti}}, \ and\ \bibinfo {author}
  {\bibfnamefont {J.~M.}\ \bibnamefont {Stafford}},\ }\href {\doibase
  10.1103/PhysRevC.101.054905} {\bibfield  {journal} {\bibinfo  {journal}
  {Phys. Rev. C}\ }\textbf {\bibinfo {volume} {101}},\ \bibinfo {pages}
  {054905} (\bibinfo {year} {2020})}\BibitemShut {NoStop}%
\bibitem [{\citenamefont {Bazavov}\ \emph
  {et~al.}(2014{\natexlab{b}})\citenamefont {Bazavov}, \citenamefont {Ding},
  \citenamefont {Hegde}, \citenamefont {Kaczmarek}, \citenamefont {Karsch},
  \citenamefont {Laermann}, \citenamefont {Maezawa}, \citenamefont {Mukherjee},
  \citenamefont {Ohno}, \citenamefont {Petreczky}, \citenamefont {Schmidt},
  \citenamefont {Sharma}, \citenamefont {Soeldner},\ and\ \citenamefont
  {Wagner}}]{Bazavov2014}%
  \BibitemOpen
  \bibfield  {author} {\bibinfo {author} {\bibfnamefont {A.}~\bibnamefont
  {Bazavov}}, \bibinfo {author} {\bibfnamefont {H.-T.}\ \bibnamefont {Ding}},
  \bibinfo {author} {\bibfnamefont {P.}~\bibnamefont {Hegde}}, \bibinfo
  {author} {\bibfnamefont {O.}~\bibnamefont {Kaczmarek}}, \bibinfo {author}
  {\bibfnamefont {F.}~\bibnamefont {Karsch}}, \bibinfo {author} {\bibfnamefont
  {E.}~\bibnamefont {Laermann}}, \bibinfo {author} {\bibfnamefont
  {Y.}~\bibnamefont {Maezawa}}, \bibinfo {author} {\bibfnamefont
  {S.}~\bibnamefont {Mukherjee}}, \bibinfo {author} {\bibfnamefont
  {H.}~\bibnamefont {Ohno}}, \bibinfo {author} {\bibfnamefont {P.}~\bibnamefont
  {Petreczky}}, \bibinfo {author} {\bibfnamefont {C.}~\bibnamefont {Schmidt}},
  \bibinfo {author} {\bibfnamefont {S.}~\bibnamefont {Sharma}}, \bibinfo
  {author} {\bibfnamefont {W.}~\bibnamefont {Soeldner}}, \ and\ \bibinfo
  {author} {\bibfnamefont {M.}~\bibnamefont {Wagner}},\ }\href {\doibase
  10.1103/PhysRevLett.113.072001} {\bibfield  {journal} {\bibinfo  {journal}
  {Phys. Rev. Lett.}\ }\textbf {\bibinfo {volume} {113}},\ \bibinfo {pages}
  {072001} (\bibinfo {year} {2014}{\natexlab{b}})}\BibitemShut {NoStop}%
\bibitem [{\citenamefont {Abelev}\ \emph
  {et~al.}(2013{\natexlab{a}})\citenamefont {Abelev} \emph {et~al.}}]{PbPb276}%
  \BibitemOpen
  \bibfield  {author} {\bibinfo {author} {\bibfnamefont {B.}~\bibnamefont
  {Abelev}} \emph {et~al.} (\bibinfo {collaboration} {ALICE Collaboration}),\
  }\href {\doibase 10.1103/PhysRevC.88.044910} {\bibfield  {journal} {\bibinfo
  {journal} {Phys. Rev. C}\ }\textbf {\bibinfo {volume} {88}},\ \bibinfo
  {pages} {044910} (\bibinfo {year} {2013}{\natexlab{a}})}\BibitemShut
  {NoStop}%
\bibitem [{\citenamefont {Abelev}\ \emph
  {et~al.}(2013{\natexlab{b}})\citenamefont {Abelev} \emph
  {et~al.}}]{PbPb276K0S}%
  \BibitemOpen
  \bibfield  {author} {\bibinfo {author} {\bibfnamefont {B.}~\bibnamefont
  {Abelev}} \emph {et~al.} (\bibinfo {collaboration} {ALICE Collaboration}),\
  }\href {\doibase 10.1103/PhysRevLett.111.222301} {\bibfield  {journal}
  {\bibinfo  {journal} {Phys. Rev. Lett.}\ }\textbf {\bibinfo {volume} {111}},\
  \bibinfo {pages} {222301} (\bibinfo {year} {2013}{\natexlab{b}})}\BibitemShut
  {NoStop}%
\bibitem [{\citenamefont {Abelev}\ \emph {et~al.}(2015)\citenamefont {Abelev}
  \emph {et~al.}}]{PbPb276phi}%
  \BibitemOpen
  \bibfield  {author} {\bibinfo {author} {\bibfnamefont {B.}~\bibnamefont
  {Abelev}} \emph {et~al.} (\bibinfo {collaboration} {ALICE Collaboration}),\
  }\href {\doibase 10.1103/PhysRevC.91.024609} {\bibfield  {journal} {\bibinfo
  {journal} {Phys. Rev. C}\ }\textbf {\bibinfo {volume} {91}},\ \bibinfo
  {pages} {024609} (\bibinfo {year} {2015})}\BibitemShut {NoStop}%
\bibitem [{\citenamefont {Abelev}\ \emph {et~al.}(2014)\citenamefont {Abelev}
  \emph {et~al.}}]{PbPb276MultiS}%
  \BibitemOpen
  \bibfield  {author} {\bibinfo {author} {\bibfnamefont {B.}~\bibnamefont
  {Abelev}} \emph {et~al.} (\bibinfo {collaboration} {ALICE Collaboration}),\
  }\href {\doibase https://doi.org/10.1016/j.physletb.2013.11.048} {\bibfield
  {journal} {\bibinfo  {journal} {Physics Letters B}\ }\textbf {\bibinfo
  {volume} {728}},\ \bibinfo {pages} {216 } (\bibinfo {year}
  {2014})}\BibitemShut {NoStop}%
\bibitem [{\citenamefont {Bellini}(2019)}]{BELLINI2019427}%
  \BibitemOpen
  \bibfield  {author} {\bibinfo {author} {\bibfnamefont {F.}~\bibnamefont
  {Bellini}} (\bibinfo {collaboration} {ALICE Collaboration}),\ }\href
  {\doibase https://doi.org/10.1016/j.nuclphysa.2018.09.082} {\bibfield
  {journal} {\bibinfo  {journal} {Nuclear Physics A}\ }\textbf {\bibinfo
  {volume} {982}},\ \bibinfo {pages} {427 } (\bibinfo {year}
  {2019})}\BibitemShut {NoStop}%
\bibitem [{\citenamefont {Adam}\ \emph {et~al.}(2020)\citenamefont {Adam} \emph
  {et~al.}}]{STAR_StrangenessPRC}%
  \BibitemOpen
  \bibfield  {author} {\bibinfo {author} {\bibfnamefont {J.}~\bibnamefont
  {Adam}} \emph {et~al.} (\bibinfo {collaboration} {STAR Collaboration}),\
  }\href {\doibase 10.1103/PhysRevC.102.034909} {\bibfield  {journal} {\bibinfo
   {journal} {Phys. Rev. C}\ }\textbf {\bibinfo {volume} {102}},\ \bibinfo
  {pages} {034909} (\bibinfo {year} {2020})}\BibitemShut {NoStop}%
\bibitem [{\citenamefont {Abelev}\ \emph {et~al.}(2009)\citenamefont {Abelev}
  \emph {et~al.}}]{STARAuAu}%
  \BibitemOpen
  \bibfield  {author} {\bibinfo {author} {\bibfnamefont {B.~I.}\ \bibnamefont
  {Abelev}} \emph {et~al.} (\bibinfo {collaboration} {STAR Collaboration}),\
  }\href {\doibase 10.1103/PhysRevC.79.034909} {\bibfield  {journal} {\bibinfo
  {journal} {Phys. Rev. C}\ }\textbf {\bibinfo {volume} {79}},\ \bibinfo
  {pages} {034909} (\bibinfo {year} {2009})}\BibitemShut {NoStop}%
\bibitem [{\citenamefont {Aggarwal}\ \emph {et~al.}(2011)\citenamefont
  {Aggarwal} \emph {et~al.}}]{STAR624}%
  \BibitemOpen
  \bibfield  {author} {\bibinfo {author} {\bibfnamefont {M.~M.}\ \bibnamefont
  {Aggarwal}} \emph {et~al.} (\bibinfo {collaboration} {STAR Collaboration}),\
  }\href {\doibase 10.1103/PhysRevC.83.024901} {\bibfield  {journal} {\bibinfo
  {journal} {Phys. Rev. C}\ }\textbf {\bibinfo {volume} {83}},\ \bibinfo
  {pages} {024901} (\bibinfo {year} {2011})}\BibitemShut {NoStop}%
\bibitem [{\citenamefont {Agakishiev}\ \emph {et~al.}(2012)\citenamefont
  {Agakishiev} \emph {et~al.}}]{STAR200}%
  \BibitemOpen
  \bibfield  {author} {\bibinfo {author} {\bibfnamefont {G.}~\bibnamefont
  {Agakishiev}} \emph {et~al.} (\bibinfo {collaboration} {STAR
  Collaboration}),\ }\href {\doibase 10.1103/PhysRevLett.108.072301} {\bibfield
   {journal} {\bibinfo  {journal} {Phys. Rev. Lett.}\ }\textbf {\bibinfo
  {volume} {108}},\ \bibinfo {pages} {072301} (\bibinfo {year}
  {2012})}\BibitemShut {NoStop}%
\bibitem [{\citenamefont {Adams}\ \emph {et~al.}(2007)\citenamefont {Adams}
  \emph {et~al.}}]{STAR200Hyperons}%
  \BibitemOpen
  \bibfield  {author} {\bibinfo {author} {\bibfnamefont {J.}~\bibnamefont
  {Adams}} \emph {et~al.} (\bibinfo {collaboration} {STAR Collaboration}),\
  }\href {\doibase 10.1103/PhysRevLett.98.062301} {\bibfield  {journal}
  {\bibinfo  {journal} {Phys. Rev. Lett.}\ }\textbf {\bibinfo {volume} {98}},\
  \bibinfo {pages} {062301} (\bibinfo {year} {2007})}\BibitemShut {NoStop}%
\bibitem [{\citenamefont {Andronic}\ \emph {et~al.}(2006)\citenamefont
  {Andronic}, \citenamefont {Braun-Munzinger},\ and\ \citenamefont
  {Stachel}}]{Feedown}%
  \BibitemOpen
  \bibfield  {author} {\bibinfo {author} {\bibfnamefont {A.}~\bibnamefont
  {Andronic}}, \bibinfo {author} {\bibfnamefont {P.}~\bibnamefont
  {Braun-Munzinger}}, \ and\ \bibinfo {author} {\bibfnamefont {J.}~\bibnamefont
  {Stachel}},\ }\href {\doibase
  https://doi.org/10.1016/j.nuclphysa.2006.03.012} {\bibfield  {journal}
  {\bibinfo  {journal} {Nuclear Physics A}\ }\textbf {\bibinfo {volume}
  {772}},\ \bibinfo {pages} {167 } (\bibinfo {year} {2006})}\BibitemShut
  {NoStop}%
\bibitem [{\citenamefont {Magestro}(2002)}]{Magestro_2002}%
  \BibitemOpen
  \bibfield  {author} {\bibinfo {author} {\bibfnamefont {D.}~\bibnamefont
  {Magestro}},\ }\href {\doibase 10.1088/0954-3899/28/7/328} {\bibfield
  {journal} {\bibinfo  {journal} {Journal of Physics G: Nuclear and Particle
  Physics}\ }\textbf {\bibinfo {volume} {28}},\ \bibinfo {pages} {1745}
  (\bibinfo {year} {2002})}\BibitemShut {NoStop}%
\bibitem [{\citenamefont {Borsanyi}\ \emph {et~al.}(2020)\citenamefont
  {Borsanyi}, \citenamefont {Fodor}, \citenamefont {Guenther}, \citenamefont
  {Kara}, \citenamefont {Katz}, \citenamefont {Parotto}, \citenamefont
  {Pasztor}, \citenamefont {Ratti},\ and\ \citenamefont
  {Szab\'o}}]{WBJuly2020}%
  \BibitemOpen
  \bibfield  {author} {\bibinfo {author} {\bibfnamefont {S.}~\bibnamefont
  {Borsanyi}}, \bibinfo {author} {\bibfnamefont {Z.}~\bibnamefont {Fodor}},
  \bibinfo {author} {\bibfnamefont {J.~N.}\ \bibnamefont {Guenther}}, \bibinfo
  {author} {\bibfnamefont {R.}~\bibnamefont {Kara}}, \bibinfo {author}
  {\bibfnamefont {S.~D.}\ \bibnamefont {Katz}}, \bibinfo {author}
  {\bibfnamefont {P.}~\bibnamefont {Parotto}}, \bibinfo {author} {\bibfnamefont
  {A.}~\bibnamefont {Pasztor}}, \bibinfo {author} {\bibfnamefont
  {C.}~\bibnamefont {Ratti}}, \ and\ \bibinfo {author} {\bibfnamefont {K.~K.}\
  \bibnamefont {Szab\'o}},\ }\href {\doibase 10.1103/PhysRevLett.125.052001}
  {\bibfield  {journal} {\bibinfo  {journal} {Phys. Rev. Lett.}\ }\textbf
  {\bibinfo {volume} {125}},\ \bibinfo {pages} {052001} (\bibinfo {year}
  {2020})}\BibitemShut {NoStop}%
\bibitem [{\citenamefont {Alba}\ \emph {et~al.}(2014)\citenamefont {Alba},
  \citenamefont {Alberico}, \citenamefont {Bellwied}, \citenamefont {Bluhm},
  \citenamefont {Sarti}, \citenamefont {Nahrgang},\ and\ \citenamefont
  {Ratti}}]{AlbaNetQ}%
  \BibitemOpen
  \bibfield  {author} {\bibinfo {author} {\bibfnamefont {P.}~\bibnamefont
  {Alba}}, \bibinfo {author} {\bibfnamefont {W.}~\bibnamefont {Alberico}},
  \bibinfo {author} {\bibfnamefont {R.}~\bibnamefont {Bellwied}}, \bibinfo
  {author} {\bibfnamefont {M.}~\bibnamefont {Bluhm}}, \bibinfo {author}
  {\bibfnamefont {V.~M.}\ \bibnamefont {Sarti}}, \bibinfo {author}
  {\bibfnamefont {M.}~\bibnamefont {Nahrgang}}, \ and\ \bibinfo {author}
  {\bibfnamefont {C.}~\bibnamefont {Ratti}},\ }\href {\doibase
  https://doi.org/10.1016/j.physletb.2014.09.052} {\bibfield  {journal}
  {\bibinfo  {journal} {Physics Letters B}\ }\textbf {\bibinfo {volume}
  {738}},\ \bibinfo {pages} {305 } (\bibinfo {year} {2014})}\BibitemShut
  {NoStop}%
\bibitem [{\citenamefont {Andronic}\ \emph {et~al.}(2019)\citenamefont
  {Andronic}, \citenamefont {Braun-Munzinger}, \citenamefont {Friman},
  \citenamefont {Lo}, \citenamefont {Redlich},\ and\ \citenamefont
  {Stachel}}]{Andronic:2018qqt}%
  \BibitemOpen
  \bibfield  {author} {\bibinfo {author} {\bibfnamefont {A.}~\bibnamefont
  {Andronic}}, \bibinfo {author} {\bibfnamefont {P.}~\bibnamefont
  {Braun-Munzinger}}, \bibinfo {author} {\bibfnamefont {B.}~\bibnamefont
  {Friman}}, \bibinfo {author} {\bibfnamefont {P.~M.}\ \bibnamefont {Lo}},
  \bibinfo {author} {\bibfnamefont {K.}~\bibnamefont {Redlich}}, \ and\
  \bibinfo {author} {\bibfnamefont {J.}~\bibnamefont {Stachel}},\ }\href
  {\doibase 10.1016/j.physletb.2019.03.052} {\bibfield  {journal} {\bibinfo
  {journal} {Phys. Lett.}\ }\textbf {\bibinfo {volume} {B792}},\ \bibinfo
  {pages} {304} (\bibinfo {year} {2019})}\BibitemShut {NoStop}%
\bibitem [{\citenamefont {Guenther}\ \emph {et~al.}(2017)\citenamefont
  {Guenther}, \citenamefont {Bellwied}, \citenamefont {Borsanyi}, \citenamefont
  {Fodor}, \citenamefont {Katz}, \citenamefont {Pasztor}, \citenamefont
  {Ratti},\ and\ \citenamefont {Szab\'o}}]{Gunther:2016vcp}%
  \BibitemOpen
  \bibfield  {author} {\bibinfo {author} {\bibfnamefont {J.}~\bibnamefont
  {Guenther}}, \bibinfo {author} {\bibfnamefont {R.}~\bibnamefont {Bellwied}},
  \bibinfo {author} {\bibfnamefont {S.}~\bibnamefont {Borsanyi}}, \bibinfo
  {author} {\bibfnamefont {Z.}~\bibnamefont {Fodor}}, \bibinfo {author}
  {\bibfnamefont {S.}~\bibnamefont {Katz}}, \bibinfo {author} {\bibfnamefont
  {A.}~\bibnamefont {Pasztor}}, \bibinfo {author} {\bibfnamefont
  {C.}~\bibnamefont {Ratti}}, \ and\ \bibinfo {author} {\bibfnamefont
  {K.}~\bibnamefont {Szab\'o}},\ }\href {\doibase
  10.1016/j.nuclphysa.2017.05.044} {\bibfield  {journal} {\bibinfo  {journal}
  {Nucl. Phys. A}\ }\textbf {\bibinfo {volume} {967}},\ \bibinfo {pages} {720}
  (\bibinfo {year} {2017})}\BibitemShut {NoStop}%
\end{thebibliography}%


%

\end{document}